\documentclass[sigconf]{acmart}
\usepackage[ruled]{algorithm2e}
\usepackage{textcomp}
\usepackage{makecell}
\usepackage{multirow}
\usepackage{hhline}
\usepackage{amsmath}
\newtheorem{Def}{Definition}
\newcommand{\vpara}[1]{\vspace{0.05in}\noindent \textbf{#1 }}
\newcommand{\hide}[1]{}

\AtBeginDocument{%
  \providecommand\BibTeX{{%
    \normalfont B\kern-0.5em{\scshape i\kern-0.25em b}\kern-0.8em\TeX}}}

\copyrightyear{2023} 
\acmYear{2023} 
\setcopyright{acmlicensed}\acmConference[WWW '23 Companion]{Companion Proceedings of the ACM Web Conference 2023}{April 30-May 4, 2023}{Austin, TX, USA}
\acmBooktitle{Companion Proceedings of the ACM Web Conference 2023 (WWW '23 Companion), April 30-May 4, 2023, Austin, TX, USA}
\acmPrice{15.00}
\acmDOI{10.1145/3543873.3584625}
\acmISBN{978-1-4503-9419-2/23/04}

\begin{document}

\settopmatter{authorsperrow=5}
\title{Continual Transfer Learning for Cross-Domain Click-Through Rate Prediction at Taobao}


\author{Lixin Liu}
\authornote{The first two authors contribute equally to the work.}
\affiliation{
  \institution{Alibaba Group}
  \city{Beijing}
  \country{China}}
\email{llx271805@alibaba-inc.com}

\author{Yanling Wang}
\authornotemark[1]
\affiliation{
  \institution{Renmin University of China}
  \city{Beijing}
  \country{China}}
\email{wangyanling@ruc.edu.cn}

\author{Tianming Wang}
\affiliation{
  \institution{Alibaba Group}
  \city{Hangzhou}
  \country{China}}
\email{tianming.wtm@alibaba-inc.com}

\author{Dong Guan}
\affiliation{
  \institution{Alibaba Group}
  \city{Beijing}
  \country{China}}
\email{jiexing.gd@alibaba-inc.com}

\author{Jiawei Wu}
\affiliation{
  \institution{Alibaba Group}
  \city{Hangzhou}
  \country{China}}
\email{wujiawei@alibaba-inc.com}

\author{Jingxu Chen}
\affiliation{
  \institution{Alibaba Group}
  \city{Hangzhou}
  \country{China}}
\email{xianlu@alibaba-inc.com}

\author{Rong Xiao}
\affiliation{
  \institution{Alibaba Group}
  \city{Hangzhou}
  \country{China}}
\email{xiaorong.xr@alibaba-inc.com}

\author{Wenxiang Zhu}
\affiliation{
  \institution{Alibaba Group}
  \city{Hangzhou}
  \country{China}}
\email{wenxiang.zwx@alibaba-inc.com}

\author{Fei Fang}
\affiliation{
  \institution{Alibaba Group}
  \city{Hangzhou}
  \country{China}}
\email{mingyi.ff@alibaba-inc.com}

\renewcommand{\shortauthors}{Liu and Wang, et al.}

\begin{abstract}
 We study the problem of cross-domain click-through rate (CTR) prediction for recommendation at Taobao.
 Cross-domain CTR prediction has been widely studied in recent years, while most attempts ignore the continual learning setting in industrial recommender systems. 
 In light of this, we present a necessary but less-studied problem named Continual Transfer Learning (CTL), which transfers knowledge from a time-evolving source domain to a time-evolving target domain. We propose an effective and efficient model called CTNet to perform CTR prediction under the CTL setting. The core idea behind CTNet is to treat source domain representations as external knowledge for target domain CTR prediction, such that the continually well-trained source and target domain parameters can be preserved and reused during knowledge transfer.
Extensive offline experiments and online A/B testing at Taobao demonstrate the efficiency and effectiveness of CTNet. CTNet is now fully deployed online at Taobao bringing significant improvements.
\end{abstract}




\maketitle

\section{Introduction}
\label{sec:intro}

Click-through rate (CTR) prediction~\cite{WDL,DCN,Zhou2017,Zhou2018,2020Search} which estimates the probability of a user clicking on a candidate item is a crucial task for recommender systems (RSs).
In practice, large-scale platforms like Taobao contain multiple different recommendation domains, and each domain maintains its own CTR prediction model.
Since some users and items are shared by different domains, cross-domain CTR prediction~\cite{li2020ddtcdr,hu2018conet,ouyang2020minet,Li2021Dual} is desired to transfer knowledge from the larger-scale source domain to alleviate the data sparsity issue in the target domain.

Previous efforts on cross-domain CTR prediction can be broadly categorized into joint learning methods~\cite{ma2018modeling,ouyang2020minet,li2020ddtcdr} and pre-training \& fine-tuning methods~\cite{chen2021user}.
The former jointly optimizes the source domain and the target domain objectives. Shared model
parameters like user/item embeddings establish
the connection between different domains.
However, such methods are inefficient to deploy online because large-scale source domain data is required for training. Besides, due to potential conflicts of different objectives~\cite{chen2021user,PCGrad}, joint training could induce negative impacts on the optimization of target domain models.
Alternatively, pre-training \& fine-tuning methods replace the target domain parameters with pre-trained source domain parameters and then fine-tune the target domain model. Since only target domain data and objective are used, pre-training \& fine-tuning methods are more efficient and effective.


In most real-world RSs, CTR prediction models are trained continually with user feedback data, such that the latest user interests can be captured in time. 
In view of this, it is necessary to study cross-domain CTR prediction under the setting of continual learning, named \textbf{Continual Transfer Learning (CTL)}. 
For this task, a straightforward solution is to use pre-training \& fine-tuning methods every time the source domain model is updated.
However, well-trained parameters of the target domain model are directly discarded and replaced by that of the source domain model. To recapture the lost historical knowledge, we need large-scale historical target domain data for fine-tuning, which makes CTL impractical.

To better solve this problem, we propose a simple and effective model called \textbf{Continual Transfer Network (CTNet)}.
The core idea is to preserve all the continually well-trained source domain and target domain parameters during CTL.
To achieve this, we treat the latest source domain representations as external knowledge for target domain CTR prediction and continually train the target domain parameters.
As shown in Figure~\ref{fig:architecture}, CTNet consists of a source tower, a target tower, and light-weighted adapters. The source tower is consistently initialized by the latest source domain model, while the target tower is continually trained with new target domain data. Adapters project the extracted source domain knowledge, so that it can be used by the target tower. Since CTNet preserves all the valuable well-trained target domain parameters, it only needs incremental target domain data to realize efficient CTL.



To sum up, the main contributions of this work include:
\begin{itemize}
    \item We study an important but less-studied problem --- continual transfer learning (CTL), which can be applied to several real-world industrial applications.
    \item We present CTNet, an effective and efficient CTL model for continual cross-domain CTR prediction. We provide useful industrial experience in deploying CTNet in large-scale RSs.
    \item We conduct extensive offline 
 and online experiments. CTNet shows high efficiency and superior performance compared with the SOTA methods. Since Dec. 2021, it has been fully deployed online at Taobao bringing significant improvements.
\end{itemize}

\section{Methods}
\label{sec:methods}

\subsection{Single-Domain Models at Taobao}
\label{sec:arch}

The large-scale e-commerce platform Taobao has multiple recommendation domains, and each of them maintains its own CTR prediction model. 
Each model aims to predict the probability of a user clicking on an item with the input of user-item features such as user features, item features, and cross features.
Specifically, these features can be summarized into categorical features like user ID, user gender, item ID, category ID, etc, and numerical features like user's or item's historical statistics. Numerical features are
pre-processed with discretization~\cite{2020AutoDis} to be categorical.
In practice, these features are transformed into dense embeddings via the embedding lookup to be concatenated as the embedding $\rm\mathbf{e}_{FEAT}$. 

Users' historical behavior sequences within all domains are introduced, which has been demonstrated crucial for CTR prediction \cite{Zhou2017}. 
Concretely, target attention (TA) \cite{Zhou2017,2020BST}, and its variants SIM (hard) \cite{2020Search} and ETA \cite{ETA} for long-term sequences are adopted to model users' behavior sequences.  
Finally, the feature embedding and the attention layer output are concatenated as $\mathbf{e} = [\rm\mathbf{e}_{FEAT}$, $\rm\mathbf{e}_{TA}$,  $\rm\mathbf{e}_{SIM}$, $\rm\mathbf{e}_{ETA}]$ to be fed into the MLP-based CTR prediction model.

\subsection{CTNet: Continual Transfer Network}
\label{sec:adapter}
\begin{figure}[t] 
\includegraphics[width=8.5cm]{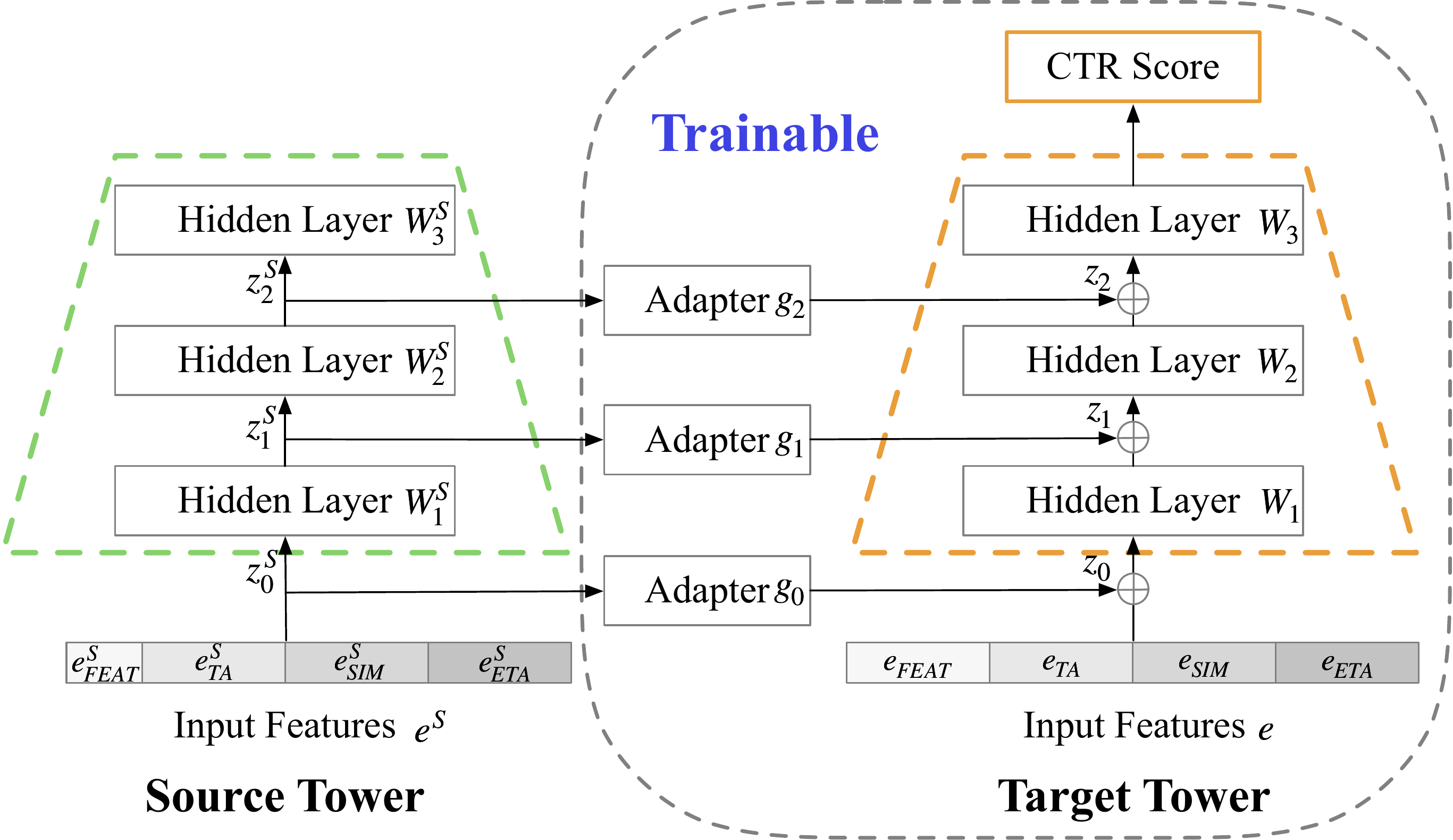}
\caption{Architecture of the proposed CTNet.}
\label{fig:architecture}
\end{figure}
Conventionally, during a time period $t \rightarrow t$+1, each single-domain model is independently trained with new user feedback data via offline incremental learning or online learning. In this work, we propose to perform continual transfer learning (CTL) and design a CTR prediction model called CTNet under the setting of CTL.

\begin{Def} \textbf{Continual Transfer Learning (CTL)}: Given a time-evolving source domain $\{\mathcal{D}^S_t\}_{t=1}^T$ and a time-evolving target domain $\{\mathcal{D}_t\}_{t=1}^T$, continual transfer learning aims to improve prediction performance on target domain $\mathcal{D}_{T+1}$ using the historical and real-time knowledge from both the source and the target domains.
\end{Def}

\vpara{Model Architecture of CTNet.}  CTNet extends the previous target domain model to be a two-tower architecture, including a source tower and a target tower.  The two towers are connected by light-weighted adapters. Figure~\ref{fig:architecture} shows an illustration of CTNet.

Specifically, the source tower has the same model architecture as the source domain model, and the target tower has the same architecture as the previous target domain model. Every time we train the CTNet, all the parameters 
of the latest source domain model, including embedding layers, attention layers, and MLP layers, are built into the computational graph of the source tower.
Since we use source domain knowledge as external information, parameters of the source tower are kept frozen during backpropagation. 

Light-weighted adapters are used as layer-wise connections of the two towers to enable domain adaptation.
Each adapter $g\left(\cdot\right)$ projects the hidden representations of source tower to be external knowledge for the target tower. Formally, the hidden representation $\mathbf{z}_{l}$ output by the $l$-th layer of target tower is computed by,
 \begin{equation}
     \mathbf{z}_{l} = \psi \left(\mathbf{W}_{l} \mathbf{z}_{l-1} + g_{l} \left(\mathbf{z}^S_{l}\right) \right), \quad l > 0
     \label{eq:add}
 \end{equation}
  \begin{equation}
     \mathbf{z}_{0} = \mathbf{e} + g_{0} \left(\mathbf{e}^S\right), \quad l = 0
     \label{eq:add2}
 \end{equation}
\noindent where $\psi$ denotes the activation function, $\mathbf{W}_l$ is the trainable weight matrix of the $l$-th layer of the target tower, and $\mathbf{z}^{S}_{l}$ is the hidden representation output by the $l$-th layer of source tower. 

We implement each adapter with a gated linear unit (GLU)~\cite{GLU} to adaptively control the information flow from the source tower to the target tower,
 \begin{equation}
     g_{l} \left(\mathbf{z}^S_{l}\right) = \mathbf{U}^1_l \mathbf{z}^S_{l} \odot \sigma \left(\mathbf{U}^2_l \mathbf{z}^S_{l}\right) 
 \end{equation}
\noindent where $\sigma$ denotes the sigmoid activation function, $\odot$ denotes element-wise vector multiplication, and $\mathbf{U}_l^{1}$ and $\mathbf{U}_l^{2}$ are trainable matrices. 

We use the output from the target tower for prediction. 
The training objective is a binary cross-entropy loss with clicked samples as positives and viewed but not clicked samples as negatives. 

\subsection{Deployment of CTNet}\label{sec:CTNet_deployment}
\begin{figure}[t] 
\includegraphics[width=8.5cm]{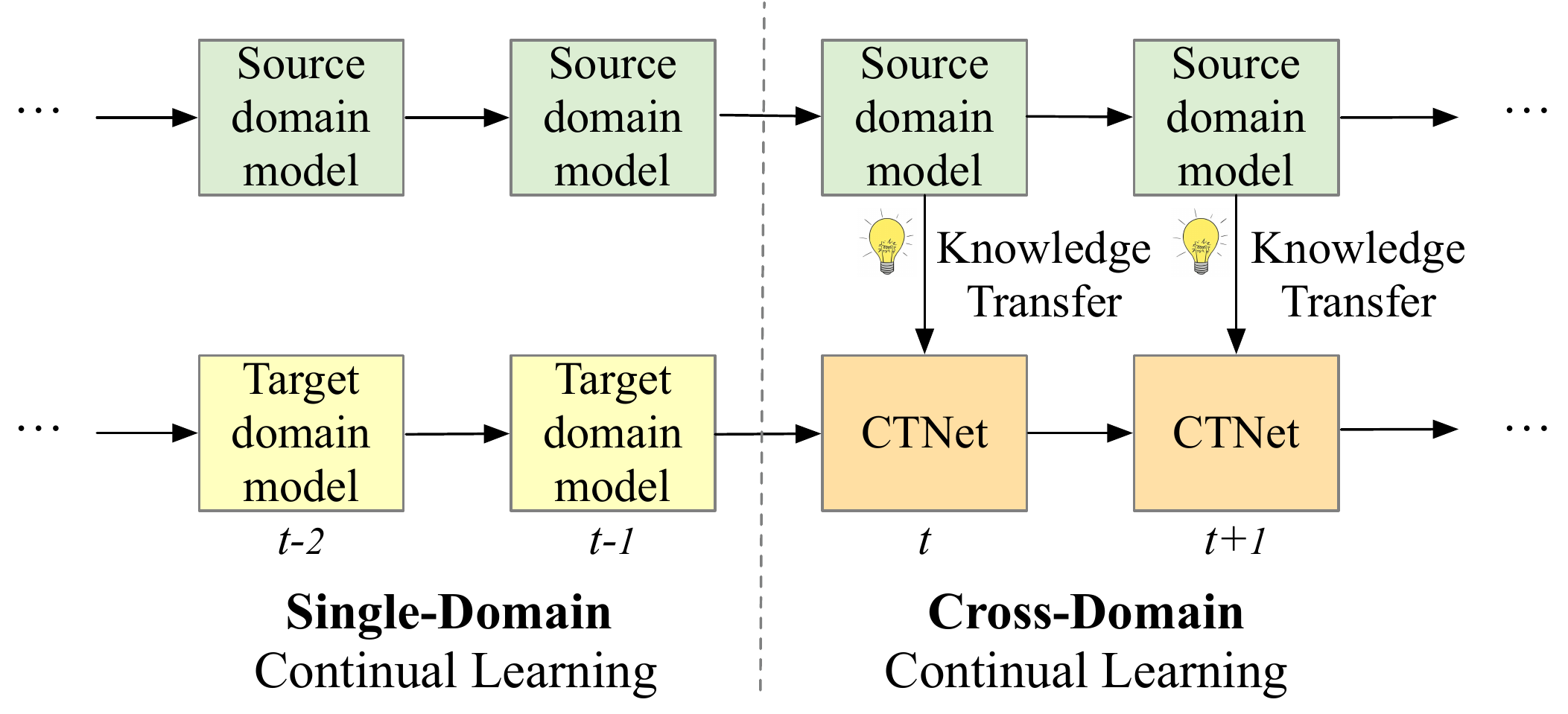}
\caption{An illustration of CTNet deployment.}
\label{fig:continual_knowledge_transfer}
\end{figure}

In this section, we provide the experience of how to deploy CTNet. To facilitate understanding, we illustrate single-domain continual learning and cross-domain continual learning in Figure~\ref{fig:continual_knowledge_transfer}. Before deploying CTNet, each single-domain model is independently and continually trained with new feedback data.
When we deploy CTNet at the time step $t$, its target tower is initialized by the latest target domain model.
More details in CTNet include,

\textbf{Two-tower model design for deriving fine-grained and real-time source domain knowledge.} \label{sec:warm}
To use the source domain knowledge, we can cache the source domain feature embeddings. 
However, such fixed knowledge cannot well fit the time-evolving target domain user behaviors and does not encode fine-grained knowledge of user-item interactions. Instead of caching  embeddings, we present the entire online two-tower architecture, which enables to model real-time user behaviors with source domain parameters.
This design will not bring extra inference latency with parallel computing, since the source domain model is lighter and runs faster than the target domain model\footnote{Source domain model has fewer input features than target domain model.}. 

\textbf{Warm start for better preserving the well-trained target domain parameters.}
We initialize the target tower of CTNet with previous target domain model. First, the adapters integrate source domain hidden representations into target domain with addition operations instead of concatenation, so that the target tower and the previous target domain model have the same architecture.
Second, the previous target domain model is highly optimized with billions of data to converge into a local optimum. Therefore, significant change of parameters makes the model deviate from this optimal solution thus hurting the performance.
Therefore, we initialize the parameters of adapters with very small values, such that the initial output of the target tower is almost identical to that of the well-trained target domain model.

\textbf{Special designs for time efficiency.} We use layer-wise adapters rather than cross-layer adapters, since this design is convenient for achieving parallel computing. 
To further reduce the computing cost, behavior sequence-based features are shared by both source tower and target tower during training and inference\footnote{
User behavior sequences encoded by TA, SIM, and ETA are from all domains in Taobao. So we can set $(\rm \mathbf{e}_{TA}, \mathbf{e}_{SIM}, \mathbf{e}_{ETA}) = (\rm \mathbf{e}^S_{TA}, \mathbf{e}^S_{SIM}, \mathbf{e}^S_{ETA})$. 
TA, SIM, and ETA modules consume the most online computing costs in our production models. Setting $(\rm \mathbf{e}_{TA}, \mathbf{e}_{SIM}, \mathbf{e}_{ETA}) = (\rm \mathbf{e}^S_{TA}, \mathbf{e}^S_{SIM}, \mathbf{e}^S_{ETA})$ will not bring negative impacts to model performances but reduce most extra computing costs.}. 
We observe that CTNet does not lead to extra online response time compared with previous target domain production models.

\section{Experiments}
\label{sec:experiments}

\subsection{Experimental Details}

\vpara{Datasets.}
Considering that there exists no suitable public benchmarks for evaluating continual cross-domain CTR prediction, we evaluate our approach on Taobao production data. Three different-sized recommendation domains --- A, B, and C are used for evaluating models' transfer performances from A to B, and A to C. Domains share some users and items, whereas domain A has covered most users and items in Taobao. We collect and sample traffic logs of these domains to get user-item interaction data within 31 days. Data from the last day is used for testing, and data from the other days is used for training. We organize datasets into periods according to the time of interactions. Each period lasts for 6 days so each dataset is split into 5 periods. The sizes of datasets in domains A, B, and C are about 150 billion, 2 billion, and 1 billion, respectively.  


\vpara{Baselines.} Various methods are compared,
including pre-training \& fine-tuning methods like
\textbf{Finetune (Embeddings)}\footnote{Update the target domain embeddings with pre-trained source embeddings and fine-tune the model on the incremental target domain data. }, \textbf{Finetune (All)}\footnote{Fine-tune the entire source domain model on the incremental target domain data.}, and 
    \textbf{Extra Embedding}\footnote{Embeddings from the latest source domain model are cached and added into the target domain model as auxiliary features. This method is similar to the online version of KEEP~\cite{keep}.
}~\cite{keep}, and joint learning methods including \textbf{MLP++}~\cite{hu2018conet},  
 \textbf{Share Bottom}~\cite{ma2018modeling}, 
 \textbf{PLE}~\cite{2020Progressive}, 
 \textbf{MiNet}~\cite{ouyang2020minet}, 
 \textbf{DDTCDR}~\cite{li2020ddtcdr}, and \textbf{DASL}~\cite{Li2021Dual}. We also report the results of single-domain models including \textbf{Base (w/o Transfer)} and \textbf{Source Model}.

\vpara{Settings.} All experiments are conducted on production models to ensure the fairness of the experiments. 
All the single-domain models have been continually trained on Taobao data for 3 years. 
For all the methods, we set the batch size to 512 and the hidden dimension of the 3-layer MLP to $[1024, 512, 256]$. 
Multi-head target attention with 8 heads and 256 hidden units is used in sequence modeling. The sub-sequence length of SIM and ETA is set to be 64. AdaGrad optimizer with an initial learning rate of $0.01$ is used for optimization. 
Following previous work in CTR prediction, we apply AUC and Group AUC (GAUC)~\cite{GAUC} for evaluation. GAUC is user group AUC weighted by the impressions of each user, which has been demonstrated to be consistent with online performance~\cite{Zhou2017}. 

\subsection{Offline Experimental Results}
\label{sec:overall_results}
 As shown in Table~\ref{tab:result}, CTNet outperforms all the baselines on the large-scale industrial dataset. More specifically,
 
(1) CTNet performs better than all the single-domain models, which demonstrates the necessity of cross-domain CTR prediction.
 
(2) Compared with pre-training \& fine-tuning methods, CTNet re-uses all the well-trained target domain parameters to minimize the loss of information. Notably, Finetune (All) discards all the well-trained target domain parameters thereby performing worse than Finetune (Embeddings), which further demonstrates the importance of preserving well-trained target domain parameters.  

(3) CTNet outperforms Extra Embedding as CTNet captures more fine-grained and real-time source domain knowledge.

(4) All the joint learning methods have relatively weak performance on the Taobao industrial dataset. These approaches transfer knowledge mainly by updating low-level features such as user/item embeddings, while the high-level representations of user-item interactions are ignored. Additionally, joint learning methods only leverage the limited 30-day source domain training data without inheriting the power of the pre-trained source domain model. Conversely, CTNet uses long-term knowledge from both the source domain and the target domain.

(5) We conduct an ablation study to show that GLU-based adapters perform better than linear layer-based adapters (CTNet (w/o GLU)), verifying the effectiveness of feature selection of gated units.

(6) We show the performance of CTNet under two different settings: continual transfer and one-time transfer. As shown in Table \ref{tab:transfer_once}, continual transfer is consistently better than one-time transfer. And the performance gain of one-time transfer drops with time. That indicates the necessity of CTL. 

\subsection{Production Deployment at Taobao}
\label{sec:production}
We deploy CTNet on two large-scale RSs at Taobao (domain B and domain C). Compared with our previous production models, CTNet respectively yielded 1.0\% and 3.6\% GAUC improvement in domain B and C (as is shown in Table \ref{tab:result}). Although domain B and C are smaller than domain A, they still have very large traffic with hundreds of millions of active users. 
For the production models, a gain of 0.1\% on GAUC is considered a significant improvement. Therefore, the performance gains from CTNet are remarkable.

We also observe significant performance gains of CTNet through online A/B testing. CTNet derives a 2.5\% CTR gain and 7.7\% Gross Merchandise Volume (GMV) gain in domain B, meanwhile derives 12.3\% CTR gain and 31.9\% GMV gain in domain C. Compared with previous production models, the offline training and online inference time does not increase with the same computational resources. Since December 2021, CTNet has been deployed fully online and serves the main traffic at Taobao. 

\begin{table}
  \centering
  \caption{Offline model evaluation results.}
    \label{tab:result}
  \setlength{\tabcolsep}{1.3mm}{
    \begin{tabular}{lcc|cc}
    \toprule
    & \multicolumn{2}{c|}{Domain A to B} &  \multicolumn{2}{c}{Domain A to C} \\   
    \hline
   Model         & AUC & GAUC & AUC & GAUC \\
\hline
\textbf{Base (w/o Transfer)} & 0.7404 & 0.6788 & 0.7104 & 0.6677 \\
\hline
\textbf{Source Model} & 0.6640 & 0.6200 & 0.5783 & 0.5771 \\
\textbf{Finetune (Embeddings) } & 0.7439 & 0.6828 & 0.7190 & 0.6769 \\
\textbf{Finetune (All)}& 0.7368 & 0.6739 & 0.6727 & 0.6352 \\
\textbf{Extra Embedding} & 0.7402 & 0.6787 & 0.7137 & 0.6693 \\
\textbf{MLP++}& 0.7405 & 0.6764 & 0.7116 & 0.6710 \\
\textbf{Share Bottom}& 0.7417 & 0.6777 & 0.7119 & 0.6726 \\
\textbf{PLE}& 0.7395 & 0.6749 & 0.7103 & 0.6679 \\
\textbf{MiNet}& 0.7411 & 0.6765 & 0.7129 & 0.6712 \\
\textbf{DDTCDR}& 0.7408 & 0.6768 & 0.7118 & 0.6709 \\
\textbf{DASL}& 0.7406 & 0.6763 & 0.7132 & 0.6714 \\
\hline
\textbf{CTNet (w/o GLU)} & 0.7465 & 0.6877 & 0.7432 & 0.7023 \\
\textbf{CTNet}& \textbf{0.7474} & \textbf{0.6888} & \textbf{0.7451
} & \textbf{0.7040}\\
    \bottomrule
    \end{tabular}
    }%
\end{table}%

\begin{table}[t]
\begin{center}
\caption{The comparisons between one-time transfer and continual transfer. } 
\label{tab:transfer_once}
\setlength {\tabcolsep} {1.2mm}
\begin{tabular}{c|ccc}
  \hline
  \toprule
  Time & Model & AUC & GAUC\\ 
  \hline
  \multirow{3}{*}{$t+\Delta t$} & \textbf{Base} (w/o transfer)& 0.7483 & 0.6931  \\
  & \textbf{CTNet} (one-time)& \textbf{0.7548} (+0.65\%) & \textbf{0.7022} (+0.91\%)  \\
  & \textbf{CTNet} (continual)& \textbf{0.7548} (+0.65\%) & \textbf{0.7022} (+0.91\%)  \\
  \hline
  
  
  \multirow{3}{*}{$t+2\Delta t$} & \textbf{Base} (w/o transfer) & 0.7416 & 0.6788 \\
  & \textbf{CTNet} (one-time) & 0.7462 (+0.46\%) & 0.6856 (+0.68\%)  \\
  & \textbf{CTNet} (continual)& \textbf{0.7486} (+0.70\%) & \textbf{0.6882} (+0.94\%)   \\
  \hline
  
  
  \multirow{3}{*}{$t+3\Delta t$} & \textbf{Base} (w/o transfer) & 0.7404 & 0.6788  \\
  & \textbf{CTNet} (one-time)& 0.7436 (+0.32\%)  & 0.6861 (+0.73\%)  \\
  & \textbf{CTNet} (continual)& \textbf{0.7474} (+0.70\%) & \textbf{0.6888} (+1.00\%)   \\
  \hline
\end{tabular}
\end{center}
\end{table}

\section{Related Work}
\label{sec:related}

\vpara{Click-Through Rate (CTR) Prediction.}
In the era of deep learning, a variety of powerful models are proposed for CTR prediction~\cite{Zhang2021,WDL,deepFM, DCN, Zhou2017,Zhou2018,2020BST,2020Search,ETA}. 
This work studies cross-domain CTR prediction~\cite{ouyang2020minet,li2020ddtcdr} which improves the performance of CTR prediction by transferring knowledge from large domains to small domains. 
A concurrent work KEEP~\cite{keep} shares similar architectures to CTNet. The main differences between KEEP and CTNet include: (1) KEEP pre-trains a concise model with data from source domain to extract the source domain knowledge.
Instead, CTNet extracts knowledge directly based on the existing source domain model, so CTNet saves computation and storage resources of pre-training and can extract more expressive knowledge from the well-trained source domain model.
(2) KEEP requests the cached source domain user-level features, item-level features, and coarse-grained interaction features (i.e., user-category interaction features) for online inference. Since the knowledge extraction and the online inference of KEEP are separated, a synchronization strategy is required to guarantee the consistency between them. Conversely, CTNet builds the parameters of the latest source domain model into the computational graph. Thus, CTNet does not cache the source domain knowledge, thereby avoiding extra synchronization strategy.
To sum up, the deployment of CTNet is easier than KEEP, and CTNet will not bring extra online latency compared to the previous production model.
Moreover, the online CTNet can extract the fine-grained and real-time knowledge of source domain user-item interactions.

\vpara{Continual Learning.}
Continual learning~\cite{2019Continual,EWC,ProgressiveNN}, also called lifelong learning, aims to learn from an infinite stream of data to gradually extend acquired knowledge for future learning~\cite{2019Continual}. 
 Studies in this field mainly focus on solving the catastrophic forgetting problem, i.e., performance on old tasks significantly degrades with the training of new tasks. However, we do not focus on this problem, since RSs are designed to predict users' future interests.

Continual learning for the cross-domain setting has not been fully studied in both academia and industry. Recent attempts such as continual pre-training~\cite{sun2020ernie,jin2021lifelong} and continual domain adaptation~\cite{liu2020learning,gong2019dlow,wang2020continuously,taufique2021conda,wu2021continuous} seem similar to our work. 
However, these attempts set only one of the source domain and the target domain to be time-evolving. Conversely, both domains are time-evolving in continual transfer learning (CTL) described in this paper.

\section{Conclusions}
\label{sec:conclusion}
In this paper, we address the problem of cross-domain CTR prediction at Taobao, and present CTNet for the task. CTNet preserves all the valuable well-trained parameters, thereby only needing incremental data to realize efficient continual transfer learning. We deploy CTNet in recommendation systems at Taobao and witness significant improvements over the existing solutions in terms of system performance and online business metrics.




\balance
\bibliographystyle{ACM-Reference-Format}
\bibliography{sample-base}
\end{document}